# Do Classics Exist in Megaproject Management?

By

By

Bent Flyvbjerg[1] and J. Rodney Turner[2]

Draft 6.1, July 2017, all rights reserved

## Abstract

This special issue asks, "Do classics exist in megaproject management?" We identify three types of classics: conventional, Kuhnian, and citation classics. We find that the answer to our question depends on the definition of "classic" employed. First, "citation classics" do exist in megaproject management, and they perform remarkably well when compared to the rest of the management literature. A preliminary Top Ten of citation classics is presented. Second, there is no indication that "conventional classics" exist in megaproject management, i.e., texts recognized as definitive by a majority of experts. Third, there is also no consensus as to whether "Kuhnian classics" exist, i.e., texts with paradigmatic clout. The importance of classics seems to be accepted, however, just as work to develop, discuss, and consolidate classics is seen as essential by megaproject scholars. A set of guidelines is presented for developing classics in megaproject management research.

**What Is a Classic?**

In conventional language a classic is a written work that is generally recognized as definitive in its field by a majority of experts in that field. Kuhn (2012, first published 1962: 10) further observes about classics of science that "[t]heir achievement was sufficiently unprecedented to attract an enduring group of adherents away from competing modes of scientific activity" and that such works served for a time to define "the legitimate problems and methods of a research field for succeeding generations of practitioners [of that field]." Finally, in an age that increasingly emphasizes research impact, Garfield (1977: 5) introduced the term "citation classics" to describe highly cited publications as identified by citation indices. All three definitions of classics – conventional, Kuhnian, and citation classics – are used by the authors of the papers included in the present special issue.

Kuhn argues that classic texts are necessary for an academic field to make progress and consolidate itself. Classic texts are important because they serve as exemplars and reference points around which paradigmatic research, or what he calls "normal science," may evolve. Paradigmatic research is used to effectively teach young scholars, or others who are new to a field, "how we do things around here" in terms of which problems to focus on; what theories, methods, and data are relevant; and what the consensus is regarding what constitutes good work in that field, according to Kuhn.

However, from time to time the consensus may break down, argues Kuhn. When that happens, paradigmatic research serves as a counterpoint against which "revolutionary" research may pit itself in developing new ideas, aimed at toppling old paradigms and arriving at new ones. Classics and their ideas may here serve as key reference points in a process that leads from an established to a toppled to a new established paradigm, through "scientific revolution," in Kuhn's words. Indeed, each scientific revolution, big or small, is likely to lead to new classic texts and the abandonment of old ones.

We find Kuhn's model useful for understanding important aspects of scientific development and we agree with his view on the role of classics. However, Kuhn developed his thinking for natural science, with most of his examples taken from physics and chemistry, and he was explicit that the situation for social science, to which research on megaproject management belongs, might be different, in that "it remains an open question what parts of social science have yet acquired such paradigms at all" (Kuhn 2012: 15). In other words, the paradigmatic status of the social sciences is uncertain, according to Kuhn.

Consequently, when we called this special issue, we wondered whether, as an emerging academic field in the social sciences, megaproject management is best described as "pre-paradigmatic" or "non-paradigmatic" (Flyvbjerg 2004: 396) – in contrast to Kuhn's notion of paradigmatic research – and whether classics can be said to exist at all in megaproject management. We were also keenly aware that the frequently shifting fashions in social science and management research should generally not be mistaken for Kuhnian paradigm shifts, and that the paradigmatic status of social science may be as uncertain today as it was at the time of Kuhn.

**Why Do Megaprojects Matter?**

The McKinsey Global Institute (2013, 2016) estimates global infrastructure spending at USD 3.4 trillion per year between 2013 and 2030, or approximately four percent of total global gross domestic product, mainly delivered as large-scale projects. In 2008, *The Economist* similarly estimated infrastructure spending in emerging economies at USD 2.2 trillion annually from 2009 to 2018 calling it "the biggest investment boom in history" (*The Economist* 2008: 80). And that's just infrastructure.

If we include the many other fields where megaprojects are a main delivery model – oil and gas, mining, mineral processing plants, aerospace, defense, IT, supply chains, mega events, etc. – then a conservative estimate for the global megaproject market is USD 6-9 trillion per year, or approximately eight percent of total global gross domestic product (Flyvbjerg 2014a). Frey (2017) goes further and estimates that megaproject spending will increase to 24 percent of global gross domestic product – or approximately USD 22 trillion per year – within the next decade. For perspective, consider that this number is larger than the GDP of any nation in the world, including the USA and China. Consider, furthermore, that recently the largest megaprojects – like the Joint Strike Fighter program and China's Belt and Road project – for the first time cost above USD 1 trillion, or more than the total market capitalization of Apple, the highest valued publicly traded company in the world. The megaproject business is big by any definition of the term.

Moreover, megaprojects have proved remarkably recession proof. In fact, the downturn from 2008 helped the megaprojects business grow further by showering stimulus spending on everything from transportation infrastructure to IT. From being a fringe activity – albeit a spectacular one – mainly reserved for rich, developed nations, megaprojects have developed into a global multi-trillion-dollar business that affects all aspects of our lives, from our electricity bill to how we shop to what we do on the Internet and how we commute.

Given this state of affairs, and given the substantial economic, social, and environmental impacts of big projects (Flyvbjerg, Bruzelius, and Rothengatter 2003), never has sound academic knowledge about megaproject management been more important. We need high-quality texts to better understand the megaproject phenomenon and to inform policy, practice, and public debate in this costly and consequential area of business and government. But what are such texts and may any of them justifiably be called classics?

**A Survey of Classics in Megaproject Management**

Our initial doubts regarding the existence of classics in megaproject management were based not only on theoretical arguments about the existence or not of paradigmatic research in megaproject management. It was also based on a recent experience Flyvbjerg had, when he accepted the invitation from a publisher to edit a volume provisionally called (by the publisher) *Classics in Megaproject Planning and Management*. The purpose was to identify and publish the classic texts in the field.

In addition to his own identification of potential articles for the planned *Classics* book, Flyvbjerg decided to conduct a survey in order to balance his possible idiosyncrasies by an outside view from experts in the field. He invited 114 persons to submit their Top Ten of the most important and influential journal articles in the field of megaproject planning and management that, in their view, should be included in the book. Respondents were explicitly asked to identify the "classics" in the field, including modern classics, that is, texts that may be considered key readings in the field of megaproject planning and management, irrespective of how old, or young, these texts were. In addition to journal articles, which were given priority, respondents were also asked to identify book chapters with seminal work. Respondents were allowed to propose any work, including their own.

Those asked were mainly scholars who themselves publish in the field of megaproject planning and management. By using Google Scholar, a systematic effort was made to include in the survey anyone who may be argued to be a published and cited scholar in the field, that is, the full population of relevant scholars. A few academics working in other fields pertinent to megaproject planning and management were also asked, as was a small number of other researchers and consultants.

The response rate of the survey was 44 percent. Of these 66 percent returned a valid Top Ten (or top whichever number of publications the respondent chose to submit, ranging from one to several dozen, with only a few respondents submitting more than ten publications). This resulted in 172 different proposed classics, suggested by respondents for inclusion in the book. Characteristics of the study are shown in Table 1.

[Table 1 app. here]

**A Top Ten of Classics?**

The overall Top Ten of classics proposed by the respondents is listed in Table 2. It should be said immediately, however, that this list must be taken with a grain of salt, as will be clear from the observations below. The average number of scholarly cites for publications on the Top Ten was 390, with the two most cited publications having cites around 1,000 and the two least cited below 100.[3] Interestingly, there is no strong correlation between a publication being on the list and that publication having high cites.

[Table 2 app. here]

Even more interestingly, agreement was scant among respondents regarding what should be considered classics in megaproject planning and management. In fact, the results from the survey indicate that if one defines a "classic" in the conventional sense, mentioned above, as a written work that is generally recognized as definitive in its field by a majority of experts in that field, then there are no classics in megaproject planning and management.

For example, we would have expected works by Albert Hirschman, Peter Hall, and Peter Morris to be identified as unequivocal classics by a majority of respondents, because to us they are classics. Not so. Only one respondent voted for a work by Hirschman (1967); four voted for a work by Hall (1980); and six for works by Morris, but with the six votes spread over five publications. On that basis, Hirschman and Morris don't even make it into the Top Ten. The highest number of votes for any of the 172 publications proposed as classics was nine (no. 1 in Table 2). This is far from the majority consensus required for something to be called a classic, according to the conventional definition.

To make matters worse, it may be argued that in order to avoid sampling bias one should exclude from the ranking (a) respondents who proposed their own publications and (b) publications authored or co-authored by Flyvbjerg, because responses may be biased in his favor due to availability bias and anchoring (Tversky and Kahneman 1973, Kahneman 1992), that is, Flyvbjerg's name may be assumed to have been more immediately available in respondents' minds than the names of other authors, because he was the one contacting the respondents and asking them to participate in the survey. Similarly, results may suffer from "politeness bias," that is, respondents trying to be courteous to Flyvbjerg, consciously or not, by proposing his work, again an argument for excluding this from the reporting of results.

If one corrects the outcomes accordingly, then the publication proposed by the most respondents was put forward by only five respondents, several times less the required majority for a classic. Remarkably, the vast majority of publications proposed as a classic – 78 percent – was put forward by one and only one respondent, indicating a very large spread in views among the respondents regarding what the classics are in megaproject planning and management.

The low number of votes on any one publication means that the Top Ten in Table 2 is not particularly robust to changes in votes, that is, the casting of one or a few votes differently might change the order of the list, or even what is on it. It should be stressed, however, that the lack of robustness is not caused by a small sample size or low response rate. As mentioned, the study may be argued to cover the whole population of scholars in megaproject planning and management, and a response rate of 44 percent of this population with 66 percent valid answers cannot be considered particularly low. The lack of robustness is caused mainly by the large spread of votes, that is, by the fundamental lack of agreement among respondents regarding what the classics are in megaproject planning and management.

In sum, these results show there is little agreement among respondents as to what the classics are in megaproject planning and management. Indeed, the results call into question the very notion of classics for this field. Consequently, Flyvbjerg and his publisher decided to change the title of the planned book from *Classics in Megaproject Planning and Management* to the more humble *Planning and Managing Megaprojects: Essential Readings* (Flyvbjerg 2014b). The first title was simply not warranted by the data, if you take seriously the conventional definition of a classic.

**Key Questions about Classics**

Based on the survey described above, we made our call for papers for the special issue, providing the following list of potential topics and questions to stimulate ideas for manuscripts (Flyvbjerg 2015: 1):

- If one defines a classic as a written work that is generally recognized as definitive in its field by a majority of experts, do classics exist in the study of megaproject management?
- If classics exist in megaproject management, what are they, how did they become classics, and what are their impacts?
- If classics do not exist in megaproject management, why not and are we likely to ever see classics in this field?
- Do classics matter to megaproject management? Why or why not? Can megaproject management thrive as an academic field without classics?
- If classics are important, how do we go about developing them in megaproject management? What can we learn on this point from other academic fields?
- Is the PMI (2013) PMBOK a classic? Is it relevant to megaproject management? What is the relationship of the PMBOK to academic work? Is the PMBOK stopping academic work from becoming classics?
- In-depth case studies of one or more classics in megaproject management were welcomed, including reasons why they must be considered classics and illustrations of their impact.

*International Journal of Managing Projects in Business* previously hosted a special issue on classics in project management (vol. 5, no. 4, 2012). That issue takes the existence of classics for granted and is focused on "revisiting the classics," without presenting evidence that classics actually exist in project management (Söderlund and Geraldi 2012: 560, 562). With the questions above, we wanted to take Kuhn (2012: 15) seriously, when he observes that for the social sciences it is an "open question" whether paradigmatic research and classics exist. We wanted to make this the key question for empirical investigation, asked for megaproject management.

We received a good batch of submissions that went through *IJPM*'s standard double blind peer review, with two anonymous referees, plus reviews by the two editors, that is, four reviews in total per paper.[4] Among the submissions, four were selected for publication, presented below. A majority of submissions are case studies of works that the authors argue are, or could be, classics in megaproject management. Interestingly, even if some of the submissions problematize the existence of classics in megaproject management, none questions the importance of classics. The authors seem to take this for granted, although several authors acknowledge the emerging character of megaproject management as an academic field and talk of "emerging classic texts" (Li et al. 2017: 1) and "the making" of classics (Siemiatycki 2016: 1).

**What the Special Issue Shows**

When discussing Table 2 above, we saw there is no strong correlation between, on the one hand, a publication being considered a classic by the respondents in the survey and, on the other, this publication having a high number of citations according to Google Scholar. This seemed strange to us. Surely one would expect a classic text to be more cited than a non-classic text, or so we reasoned. We also saw above that the ranking in Table 2 lacked robustness. Clearly a proper bibliometric study of publications in megaproject management was needed, something which has not been done before. Perhaps such a study would produce more robust results, we hoped. Fortunately, two bibliometric studies were submitted for the special issue and the best of these was chosen for publication, "Bibliographic and Comparative Analyses to Explore Emerging Classic Texts in Megaproject Management" by Li et al. (2017).

This study benchmarks megaproject management against five established fields in the management literature, namely institutional theory, organizational effectiveness theory, stakeholder theory, top management theory, and resource dependence theory. Compared with these fields, which started to build their classics and citations 30 and more years ago, megaproject management research is young, with classic texts beginning to emerge only 12 to 13 years ago, according to Li et al. (2017: Figure 5). Based on bibliometric analyses of texts in the benchmark, Li et al. find, first, that it takes 20 years or more after initial publication of classic texts for these to have the surge in citations that is characteristic of such texts, and, second, that the most cited texts in megaproject management are too young to have gone through the full cycle of a classic text, including the surge. The young age of megaproject management research could possibly explain any lack of consensus regarding what the classics are in the field, or whether they even exist, according to this argument.[5]

Nevertheless, megaproject management is doing well compared with the benchmark in the sense that total citations for the most cited texts in megaproject management is higher than total citations for the five fields in the benchmark, except one, when compared for the first 10 years after the initial publication of the most cited texts in each field (Li et al. 2017: 20). Moreover, the bibliometric study finds that several texts in megaproject management published as recently as 2014 and 2015 have been listed among the coveted one percent of highly cited papers in the ESI (Essential Science Indicators, a statistical platform under Web of Science), indicating the quick growth and significant potential for impact of research in megaproject management, according to Li et al. (p. 20). These results bode well for the continued development of megaproject management as an emerging academic field and for the development of citation classics within this field.

Li et al. (2017: App. 2, 3) record two Top Ten lists of the most cited texts in megaproject management, one based on megaprojects as keyword only and one based on megaprojects and other relevant terms as keywords. The two lists are included in Table 3 together with the number of citations recorded by Li et al. for each publication. Li et al. follow Garfield (undated), who defines citation classics as follows: "In general, a publication cited more than 400 times should be considered a classic; but in some fields with fewer researchers, 100 citations might qualify a work." We see that citation classics in Garfield's sense clearly exist in megaproject management research. Comparing Table 2 and Table 3, we see that the Top Tens are quite different, although there is some overlap with three publications appearing in both tables. The tables measure different things, so differences should be expected. Table 2 counts the number of people who agreed in a survey that a text should be considered a classic, whereas Table 3 counts how many times a given text has been cited according to a bibliometric study. The rankings in Table 3 are more robust than those in Table 2, in the sense that it would take substantial changes in the citations that form the basis for Table 3 for the rankings in that table to change considerably, whereas, for Table 2, even a few changes in the votes of the respondents in the survey might change the ranking, and even the individual publications included, as we saw above. However, the rankings in Table 3 may be sensitive to the choice of keywords used to decide which publications to include in the bibliometric analyses. For instance, it seems an arbitrary consequence of choice of keywords that Hall (1980), with more than 800 citations according to Google Scholar, is not included in Table 3. Further study is needed to better understand the effect on results of the choice of keywords. We suggest this as an idea for a future article in *IJPM*.

[Table 3 app. here]

In "The Making and Impacts of a Classic Text in Megaproject Management: The Case of Cost Overrun Research," Siemiatycki (2016) argues that Flyvbjerg et al. (2002) has reached the status of a classic in the Kuhnian sense. We note that this work is also identified as a citation classic by Li et al. (2017) and appears in the Top Tens in Tables 2 and 3.

Siemiatycki teases out five "key takeaway lessons" (pp. 8-9), which he uses both to explain his conclusions regarding Flyvbjerg et al. (2002) and to proactively identify a "pathway" forward that can be followed by anyone interested in producing future classic texts in the field of megaproject management:

1. Novelty is not required to be revolutionary in Kuhn's sense.
2. Effective communication to its audience – academic or public – is a key feature of classic texts.
3. Paradigm-shifting research depends on widespread awareness of the work among relevant scholars in the field.
4. Dissemination of research in the mass media is an effective technique to increase the impact of megaproject management research on practice.
5. Authors of classic texts should be prepared for pushback from the standard bearers of the established paradigm, in academia, policy, and practice.

In "Classics in Megaproject Management: A Structured Analysis of Three Major Works," Pollack et al. (2017) assess whether Morris and Hough (1987), Merrow (2011), and Flyvbjerg et al. (2003) should be considered classics in megaproject management, again in the Kuhnian sense, supplemented by Calvino's (2000) take on classics. Pollack et al. employ four criteria in making their assessment: (a) overall impact of the text, (b) impact over time, (c) contribution to formation of a distinct discipline, and (d) personal impact on the reader, depth, and insight. Following these four criteria, the authors conclude that none of the three books under consideration are classics in megaproject management. Two of these books are identified as citation classics by Li et al. (2017).

Morris and Hough (1987) could possibly be viewed as a classic of general project management research, according to Pollack et al., but should not be seen as a classic in megaproject management research as it is not sufficiently specialized. Second, Merrow (2011) makes a significant contribution to megaproject research, but falls short of being a classic in the field, partly because the book has had insufficient time to make its impact and partly because it is unclear whether the data supporting the book, which come from a private consultancy founded and led by the author, has been subjected to independent and scholarly peer review. Finally, of the three works analyzed, Flyvbjerg et al. (2003) makes the greatest claim for the status of a classic in megaproject management research, according to Pollack et al., but this work, too, falls short, in particular as regards a need to address the internal "politicality" of megaprojects through ethnographic real-time research.

In sum, Pollack et al. find that megaproject management has good candidates for classic texts, but they are not quite there yet. In particular, they need more use of actor-network theory, which happens to be the forte of the authors: "[A] future classic must include those ideas to address some of the field's most prominent questions," conclude Pollack et al. (2017: 11).

Finally, in "The Fate of Ideals in the Real World: A Long View on Philip Selznick's Classic on the Tennessee Valley Authority," Ansar (2017) identifies an interesting paradox about Selznick's (1949) study of megaproject management with the TVA, which has likely delivered more megaprojects than any similar organization in the Americas: On one hand, Selznick's book must be regarded as a classic, by any measure – including more than 4,000 citations, according to Google Scholar – in management and organization studies; on the other, megaproject (and project) management scholars do not cite the book, even though it can be argued to be one of their founding classics, according to Ansar.

In explaining this paradox, Ansar identifies a basic ignorance in project management scholarship, including megaproject management. Scholars in project management ignore Selznick because they operate within a paradigm based in engineering and economics and are not sufficiently socialized into the broader field of management studies and social science, where sociological, political, organizational, and behavioral approaches, like those used by Selznick, are mainstream. Instead, project management scholars tend to see Selznick as an outsider producing "uncomfortable knowledge" that not only does not fit their worldview but problematizes it, and is therefore best ignored, they think, according to Ansar.

Selznick (1949) is not the only classic that has been disregarded like this and the cost to project management scholarship is immense in terms of weak thinking, when compared to other parts of academia, and poor results, when ideas are applied in practice. The solution is for project management scholars to revisit and reintegrate their ignored classics. Only by doing this can we fundamentally recast the direction of scholarship in our field, away from the narrow instrumentalism that currently dominates toward thinking that builds on the full range of social science theory, methodology, and data – as for management scholarship in general – concludes Ansar.

**The Status of Classics in Megaproject Management**

We conclude that the answer to the question of whether classics exist in megaproject management research depends in part on the definition employed of what a "classic" is. In terms of Garfield's (1977) bibliometric "citation classics," we conclude with Li et al. (2017) that based on the available evidence megaproject management is doing remarkably well. This conclusion applies to older, highly cited texts in megaproject management – like Morris and Hough (1987), Altshuler and Luberoff (2003), and Flyvbjerg et al. (2002, 2003) – which compare favorably to the most cited texts in more established fields of management research. But the conclusion also applies to more recent texts, as witnessed by new works in megaproject management that have quickly made it onto the ESI list of top one percent of highly cited papers, for instance Ansar et al. (2014) and Follmann (2015).

Furthermore, if we employ Garfield's standard threshold of 400 citations for a text to be considered a classic, then no less than ten of the publications on the two Top Ten lists in Table 3 are classics. If, instead, we use Garfield's threshold of 100 citations, which he recommends for smaller fields with fewer researchers, to which megaproject management research can easily be argued to belong, then all the Top Ten publications in Table 3 get to count as classics.[6]

In sum, the results from the first bibliometric study of megaproject management research bode well for the continued development of megaproject management as an emerging academic field and for the development of classics, in the bibliometric sense, within this field. However, before we get too excited about these results the bibliometric studies on which they are based must be replicated with more and better data, as acknowledged by Li et al. (2017: 31). In particular, more and different keywords than those initially used must be applied in further studies in order to assess the robustness of the findings.[7]

Pollack et al. (2017) is the perfect antidote to getting overexcited about Li et al. Employing the Kuhnian definition of classics – emphasizing novelty, attraction, endurance, and paradigmatic clout – Pollack et al. assess three potential classics in megaproject management, namely Morris and Hough (1987), Merrow (2011), and Flyvbjerg et al. (2003). Two of these are identified as classics by Li et al., as mentioned above. In contrast, Pollack et al. find that none of the three deserves the distinction. Judged on its own terms, this result is just as valid as that by Li at al., which illustrates well our point that conclusions about which texts qualify as classics depend on the definition of classic employed.

The three potential classics chosen by Pollack et al. may be argued to be particularly likely cases of classics in megaproject management research. Consequently, an implication of Pollack et al.'s conclusion is that if none of these three works can be argued to be classics, then we should expect that perhaps no, or only few, works in megaproject management research can be. This conclusion fits well with our own initial assumptions, based on the survey results presented above, when we first made the call for papers for this special issue. The conclusion does not fit, however, with Siemiatycki's (2016) clear identification – also based on the Kuhnian definition – of Flyvbjerg et al. (2002) as a classic in the field, nor with Ansar's (2017) conclusion that Selznick (1949) is a classic, albeit an ignored one by megaproject management scholars.

We conclude, regarding the question of classics in megaproject management research:

- Whether classics exist in this field depends on the definition of "classic" employed.

- There is no general consensus among scholars in the field regarding what the classics are, or of the definition of what a classic is.

- Classics do exist in the field in the sense of "citation classics," and these perform remarkably well when compared to other classics in the management literature, for both older and newer classics.

- There is no agreement as to whether "Kuhnian classics" exist in this field. Some scholars argue they do, others they don't. Furthermore, some argue that works identified as citation classics do not qualify as Kuhnian classics, indicating that the bar may be higher for a work to count as a Kuhnian classic than as a citation classic.

- There is no indication that "conventional classics" exist in the field, in the sense of works that are recognized as definitive by a majority of experts. No majority and no consensus have been documented on this point, but quite the opposite.

- Works exist for which it may be argued convincingly that they ought to be considered classics in the field, but they are not, because scholars in the field ignore them, to the detriment of the development of the field.

**The Way Forward for Classics in Megaproject Management**

Going forward, we see the following main tasks to stimulate and improve scholarship in megaproject management, in terms of classics:

- *The classics that have emerged over the past ten to 15 years, new and old, should be further developed and consolidated*. We are pleasantly surprised to see how strong megaproject management research stands when compared to other areas of management research, in terms of citations. This is cause for celebration – but not for resting on our laurels. It takes 20 to 30 years to grow and consolidate a classic, so for megaproject management this work has only just begun, with the most important part to be done over the next ten to 15 years. This work needs serious attention, including the updating of emerging classics when and if new data and theories warrant this, as suggested by Li et al. (2017).

- *New classics should be developed*, focused on high-quality data and using the strongest theories from across the social sciences, for instance following the guidelines set out by Siemiatycki (2016) and listed above. Too much scholarship in megaproject and project management build on weak, idiosyncratic data and weak theory with little impact. This should be avoided going forward by learning from best practice in other parts of the social sciences, and by emphasizing impact: in the academy, policy, practice, and civil society. More publication should be in high-quality journals and with high-quality academic publishers at the best universities, to secure vetting and quality control that would increase the likelihood of classics emerging from the process.

- *Ignored classics should be revisited and taken seriously*, including where such texts are critical of megaproject management. To ignore criticism because it constitutes "uncomfortable knowledge" is the road to mediocrity, or worse. Criticism is the lifeblood of scholarship and should be welcomed. Ansar (2017) points to Selznick (1949) as an ignored classic, but others come to mind, including Hirschman (1967) and Caro (1974). Söderlund and Geraldi (2012) similarly point to Sayles and Chandler (1971) and Sapolsky (1972). Megaproject management is not in a position where it can afford to overlook important contributions to its knowledge base. These works, authored by people who have thought long and hard about megaproject management before most others, need to be carefully considered and integrated in the canon.

- *Debate should continue*, and should be encouraged, about which texts may and may not be considered classics, with full knowledge that it may be difficult, or impossible, to reach a consensus about what the classics are, or even that they exist. No field deserves its place in the academy without engaging seriously in such debate. Flyvbjerg (2014b, 2017) were developed to support this debate by identifying and publishing candidate texts for discussion.

"History suggests that the road to a firm research consensus is extraordinarily arduous," said Kuhn (2012: 15) about arriving at paradigmatic research and classics. Making progress may therefore prove difficult, in terms of agreeing on what classics and paradigmatic research may be in megaproject management. But difficulty should not keep us from trying, needless to say, which is what we have done with this special issue. We intend the issue as a beginning and encourage others to follow up. There is much to gain from the effort, as documented by the contributions that follow.

*Table 1: Characteristics of survey asking, "What is your Top Ten of classics in the field of megaproject planning and management?"*

| | |
|---|---|
| Number of people asked | 114 persons |
| Percentage that responded | 44% |
| Of these, percentage with valid and reliable answers | 66% |
| Total number of different classics proposed | 172 publications |
| Of these, the publication proposed by the most persons was proposed by | 9 persons |
| Percentage of the 172 publications with more than 1000 cites on Google Scholar | 10% |
| Percentage of the 172 publications with more than 100 cites on Google Scholar | 43% |

*Table 2: Top Ten list of "classics" in megaproject planning and management, based on survey by Flyvbjerg (2014a). If publications by Flyvbjerg were omitted, in order to avoid possible sampling bias (see main text), then Wachs (1989) and Kain (1990) would be included as number 9 and 10 on the list, respectively.*

| No.* | Authors | Title |
|---|---|---|
| 1 | Flyvbjerg, Holm, Buhl | Underestimating Costs in Public Work Projects. Error or Lie? |
| 2 | Pickrell | A Desire Named Streetcar: Fantasy and Fact in Rail Transit Planning |
| 3 | Pitsis, Clegg, Marosszeky, Rura-Polley | Constructing the Olympic Dream: A Future Perfect Strategy of Project Management |
| 4 | Hall | Great Planning Disasters |
| 5 | Miller, Lessard | The Strategic Management of Large Engineering Projects: Shaping Institutions, Risks and Governance |
| 6 | Stinchcombe, Heimer | Organization Theory and Project Management: Administering Uncertainty in Norwegian Offshore Oil |
| 7 | Kahneman, Lovallo | Timid Choices and Bold Forecasts: A Cognitive Perspective on Risk Taking |
| 8 | Flyvbjerg, Bruzelius, Rothengatter | Megaprojects and Risk: An Anatomy of Ambition |
| 9 | Lovallo, Kahneman | Delusions of Success: How Optimism Undermines Executives' Decisions |
| 10 | Miller, Lessard | Understanding and Managing Risks in Large Engineering Projects |

*) In case of a tie, the publication with the highest number of citations according to Google Scholar was listed first.

*Table 3: Top Ten lists of most cited texts in megaproject management, based on citations according to bibliometric analyses by Li et al. (2017).*

| No. | A. Megaprojects as keyword | Cites A. | B. Megaprojects and relevant terms as keywords | Cites B. |
|---|---|---|---|---|
| 1 | Megaprojects and Risk: An Anatomy of Ambition (Flyvbjerg, Bruzelius, Rothengatter) | 2061 | Megaprojects and Risk: An Anatomy of Ambition (Flyvbjerg, Bruzelius, Rothengatter) | 2061 |
| 2 | Mega-Projects: The Changing Politics of Urban Public Investment (Altshuler, Luberoff) | 576 | Underestimating Costs in Public Works Projects: Error or Lie? (Flyvbjerg, Holm, Buhl) | 982 |
| 3 | Globalization and Urban Change: Capital, Culture, and Pacific Rim Mega-Projects (Olds) | 374 | The Anatomy of Major Projects: A Study of the Reality of Project Management (Morris, Hough) | 907 |
| 4 | Globalization and the Production of New Urban Spaces: Pacific Rim Megaprojects in the Late 20th Century (Olds) | 208 | Neoliberal Urbanization in Europe: Large-Scale Urban Development Projects and the New Urban Policy (Swyngedouw, Moulaert, Rodriguez) | 627 |
| 5 | Managing Public-Private Megaprojects: Paradoxes, Complexity, and Project Design (van Marrewijk, Clegg, Pitsis, Veenswijk) | 197 | Mega-Projects: The Changing Politics of Urban Public Investment (Altshuler, Luberoff) | 576 |
| 6 | Understanding the Outcomes of Megaprojects: A Quantitative Analysis of Very Large Civilian Projects (Merrow) | 162 | Evaluating the Risks of Public Private Partnerships for Infrastructure Projects (Grimsey, Lewis) | 571 |
| 7 | Mega-Projects in New York, London and Amsterdam (Fainstein) | 150 | Causes of Delay in Large Construction Projects (Assaf, Al-Hejji) | 549 |
| 8 | Big Decisions, Big Risks. Improving Accountability in Mega Projects (Bruzelius, Flyvbjerg, Rothengatter) | 135 | How (In)Accurate Are Demand Forecasts in Public Works Projects? The Case Of Transportation (Flyvbjerg, Holm, Buhl) | 494 |
| 9 | Old Mega-Projects Newly Packaged? Waterfront Redevelopment in Toronto (Lehrer, Laidley) | 130 | The Strategic Management of Large Engineering Projects: Shaping Institutions, Risks, and Governance (Miller, Lessard) | 474 |
| 10 | Decision-Making on Mega-Projects: Cost-Benefit Analysis, Planning and Innovation (Priemus, Flyvbjerg, van Wee) | 129 | Rescuing Prometheus: Four Monumental Projects That Changed Our World (Hughes) | 467 |

# Notes

<sup>1</sup> Bent Flyvbjerg is the first BT Professor and inaugural Chair of Major Programme Management at Saïd Business School, University of Oxford. He is guest editor of this special issue of *IJPM*.

<sup>2</sup> Rodney Turner is Professor of Project Management at the SKEMA Business School, in Lille France. He is editor of *IJPM*.

<sup>3</sup> All citations are from Google Scholar and were measured December 2011, when the survey was done.

<sup>4</sup> Full disclosure: For papers that were about Flyvbjerg's publications, we decided he had a conflict of interest as guest editor and the decision on whether or not to publish these papers was left to the editor in collaboration with the referees.

<sup>5</sup> An alternative explanation for a lack of consensus would be that the field of megaproject management research is multidisciplinary and fragmented to a degree where classics are unlikely to ever appear, because a consensus on what they are is unlikely to materialize.

<sup>6</sup> In both instances, this is when using Google Scholar to count citations. If, instead, Web of Science (WOS) is used, then only one publication from Table 3 gets to count as a classic with the 400-citations threshold, but nine publications are classics with the 100-citations threshold. Given the fact that many of the potential classics in megaproject management are books, Google Scholar seems the best index for measuring citations, as WOS is known to have a bias against books and for papers.

<sup>7</sup> Although we find Garfield's definition and thresholds useful for getting a first grasp of which texts may be important, we are skeptical of their mechanical nature and of the fact that they seem to identify too many texts as classics. Discussions like those of Pollack et al. (2017), Siemiatycki (2017), and Ansar (2017) are needed to consolidate what the classics may be and to qualify a possible consensus, in our judgment.